\documentclass[pdflatex,sn-mathphys-num]{sn-jnl}

\usepackage{graphicx}%
\usepackage{multirow}%
\usepackage{amsmath,amssymb,amsfonts}%
\usepackage{amsthm}%
\usepackage{mathrsfs}%
\usepackage[title]{appendix}%
\usepackage{xcolor}%
\usepackage{textcomp}%
\usepackage{braket}
\usepackage{manyfoot}%
\usepackage{booktabs}%
\usepackage{algorithm}%
\usepackage{algorithmicx}%
\usepackage{algpseudocode}%
\usepackage{listings}%

\theoremstyle{thmstyleone}%
\theoremstyle{thmstyletwo}%

\theoremstyle{thmstylethree}%

\begin{document}

\title[Article Title]{Changing the Game: The Bounce-Bind Ising Machine}


\author[1]{\fnm{Haiyang} \sur{Zhang}}\email{HaiyangZ@whu.edu.cn}
\equalcont{These authors contributed equally to this work.}

\author*[1]{\fnm{Hao} \sur{Wang}}\email{wanghao@whu.edu.cn}
\equalcont{These authors contributed equally to this work.}

\author[1]{\fnm{Rui} \sur{Zhou}}\email{2019302020021@whu.edu.cn}

\author*[1]{\fnm{Sheng} \sur{Chang}}\email{changsheng@whu.edu.cn}

\affil*[1]{\orgdiv{School of Physics and Technology}, \orgname{Wuhan University}, \orgaddress{\city{Wuhan}, \postcode{430072}, \state{Hubei}, \country{P. R. China}}}


\abstract{
The Ising model, originally proposed a century ago, has become a cornerstone of combinatorial optimization in recent decades. However, Ising machines remain constrained by a fundamental hardware-speed trade-off. We introduce the Bounce-Bind Ising Machine (BBIM), a mechanism with a single tunable parameter that modulates spin dynamics without altering the energy landscape, building upon the classic golf-ball analogy but replacing it with a dynamic tennis ball/shot put system. The Bounce mode (accelerating escapes from local minima) and Bind mode (enabling rapid convergence) dynamically balance speed and quality. Benchmarked on dense MAX-CUT (edge density=0.5), BBIM achieves a peak speedup of 6.15× at n=200. For sparse 3-Regular 3-XORSAT (second-order), the peak speedup reaches 27.3× at n=160. Both results incur negligible additional hardware resource consumption. This work demonstrates a critical pathway to circumventing the hardware-speed bottleneck and its practical applicability to large-scale optimization hardware, validated on structurally distinct benchmarks.

}





\maketitle

\section{Introduction}\label{sec1}
The Ising model, first introduced in the 1920s to study phase transitions in ferromagnetic materials\cite{lenz1920beitrvsge}, underpins modern Ising machines designed to find ground-state configurations of complex energy landscapes for combinatorial optimization problems (COPs). Nowadays, Ising machines as hardware solvers based on the Ising model have gained widespread application in efficiently solving COPs\cite{mohseni2022ising}, many of which correspond to NP problems that are computationally challenging for conventional computing architectures. These machines such as D-Wave’s quantum annealing machines\cite{santoro2002theory, hauke2020perspectives}, coherent Ising machines \cite{inagaki2016coherent,wang2013coherent}, and CMOS Ising machines \cite{su2022flexspin,oku2019fully} utilize annealing algorithms and probabilistic techniques such as Gibbs sampling to achieve optimal or near-optimal solutions. For these machines, the Ising problem is described with the Hamiltonian (Ising energy):
\begin{equation}
E_0=-\sum_{i<j}J_{ij}m_im_j-\sum_ih_im_i \label{E0}
\end{equation}
where $\boldsymbol{J}$ denotes the interaction matrix, and $\boldsymbol{h}$ is the external magnetic field vector. The update function for existing Ising machines is derived from the energy gradient:
\begin{equation}
I_i=-\frac{\partial E}{\partial m_i}=\sum_{j\neq i}J_{ij}m_j+h_i  \label{Ii}
\end{equation}
\begin{equation}
m_i(t+1)=\operatorname{sign} [\tanh(\beta I_i) - \operatorname{rand}(-1,1)] \label{mi_next}
\end{equation}
where $\beta$ is the inverse temperature of the system. When the system reaches thermal equilibrium, the probabilities of each state follow the Boltzmann distribution:
\begin{equation}
    P(\boldsymbol{m}) = \frac{1}{Z} \exp\left(-\frac{E(\boldsymbol{m})}{T}\right), \quad Z = \sum_{\boldsymbol{m}}e^{-\frac{E(\boldsymbol{m})}{T}} \label{boltzmann_law}
\end{equation}
where $Z$ is the partition function, and $T$ is the temperature. The solution to Eq. (\ref{E0}), under the widely used analogy of the ``golf course"\cite{katzgraber2015good,dobrynin2024energy}, involves identifying the correct global minimum among many local minima. This is akin to hitting a golf ball into one specific hole out of many on a green. The key challenge for Ising machines is efficiently navigating through these local minima to reach the ground state. 

In this work, we introduce a new term to the Hamiltonian, a constant carefully designed not to alter the energy landscape but to modulate the update dynamics of the Ising Machine. This new term enables the use of a continuously tunable parameter, which effectively modulates the convergence behavior of the system. Depending on the parameter's value, the iteration exhibits two distinct modes: rapid state transitions (referred to as the ``Bounce" mode) and stable state persistence (referred to as the ``Bind" mode). Bounce-dominant behavior, characterized by rapid bouncing like a tennis ball, enables quick state flips to escape local minima. Bind-dominant behavior, akin to the inertia-driven binding of a shot put, maintains spin states to promote gradual convergence. These modes expand the Ising Machine’s versatility, enabling faster convergence and improved success rates across diverse energy landscapes. Therefore, the new term is named the Bounce-Bind (BB) term, and the modified Ising Machine is referred to as the Bounce-Bind Ising Machine (BBIM). Notably, when the BB parameter term is set to zero, it corresponds to the standard Ising Machine, i.e., it reduces to the normal golf ball in the game.

Here, we take dense MAX-CUT problems\cite{commander2009maximum} and sparse 3-Regular 3-Exclusive-OR Satisfiability (3R3X) problems\cite{kowalsky20223} as the benchmark problems. The MAX-CUT problem is a typical NP-hard problem, and it has significant advantages when using Ising machines for solution interpretation. Due to the equation planting, controllable ground state degeneracy and topological structure\cite{hen2019equation}, 3R3X problems are widely applied to evaluate the effectiveness of heuristic Ising solvers. 

In the following sections, we first introduce the Bounce-Bind mechanism and its influence on the search process of the Ising machine, using simple examples to explore the two modes. We then present the basic structure of BBIM implemented on FPGA and briefly introduce the benchmark problems. Subsequently, we compare the classical Ising machine and the BBIM in terms of success rates and solving speed on small-scale problems (sparse 3R3X and dense MAX-CUT) and large-scale problems (sparse and dense MAX-CUT). Finally, the results demonstrate that the BBIM outperforms the classical Ising machine using the same algorithm.



\section{Results}\label{sec2}

\subsection{The Bounce-Bind term and its effect on the update process}\label{bbeft}

The Hamiltonian of the Ising model with Bounce-Bind term is given by:
\begin{equation}
E_{BB}=E_0+E_\mathcal{B}=-\sum_{i<j}J_{ij}m_im_j-\sum_ih_im_i-\frac{\mathcal{B}}{2}\sum_im_i^2 \label{2nd-ising-bb}
\end{equation}
where the first term $E_0$ represents the classical Ising Hamiltonian, $E_\mathcal{B}$ is the Bounce-Bind term, and $\mathcal{B}$ denotes the continuously adjustable Bounce-Bind parameter. Because of the bi-polarization property of spin state $m_i$, the Bounce-Bind term contributes a constant value of $-\mathcal{B}/2$. As a result, the energy landscape remains equivalent to the classical one described with Eq. (\ref{E0}), ensuring the physical problem is preserved. However, the derivative function for Eq. (\ref{2nd-ising-bb}) becomes:
\begin{equation}
    I_{BB,i} = I_i+\mathcal{B}m_i
\end{equation}
where $I_i$ is calculated by Eq. (\ref{Ii}). The term $\mathcal{B}m_i$ significantly alters the behavior of the Ising Machine. Notably, Eq. (\ref{mi_next}), which defines the spin update rule, remains unchanged in this work. When $\mathcal{B}=0$, our BBIM is exactly the same as the classical Ising machine.

\begin{figure}[h]
\centering
\includegraphics[width=1.0\textwidth]{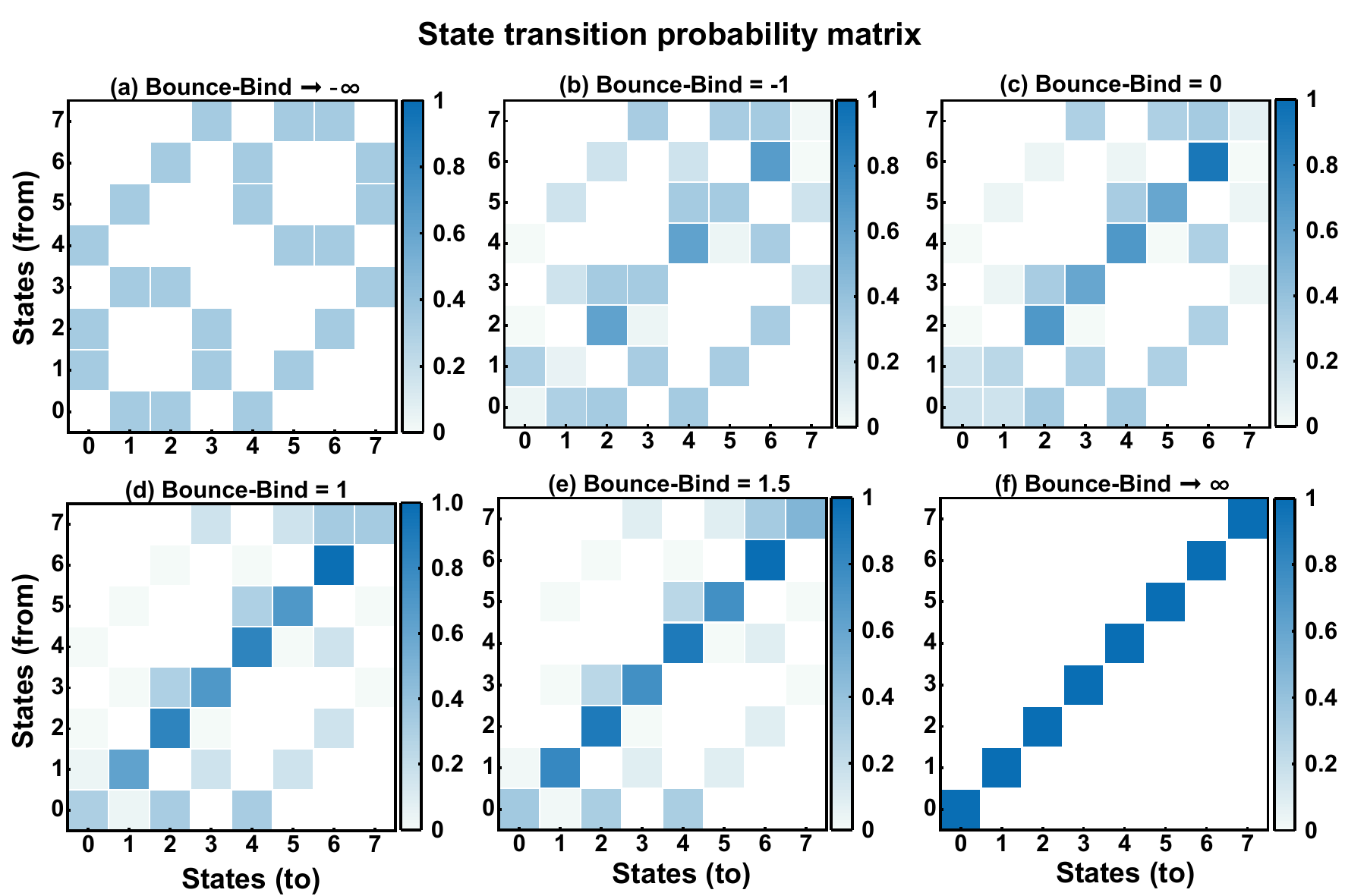}
\caption{\textbf{State transition matrices under different Bounce-Bind parameters in a case of a 3-bit Ising problem.} The results show that the probability of the system transitioning to its own state increases monotonically with the parameter $\mathcal{B}$, which influences the convergence of the system. The state transition probability is calculated by the algorithm in Supplementary Section S1.}\label{stp}
\end{figure}

\begin{figure}[h]
\centering
\includegraphics[width=1.0\textwidth]{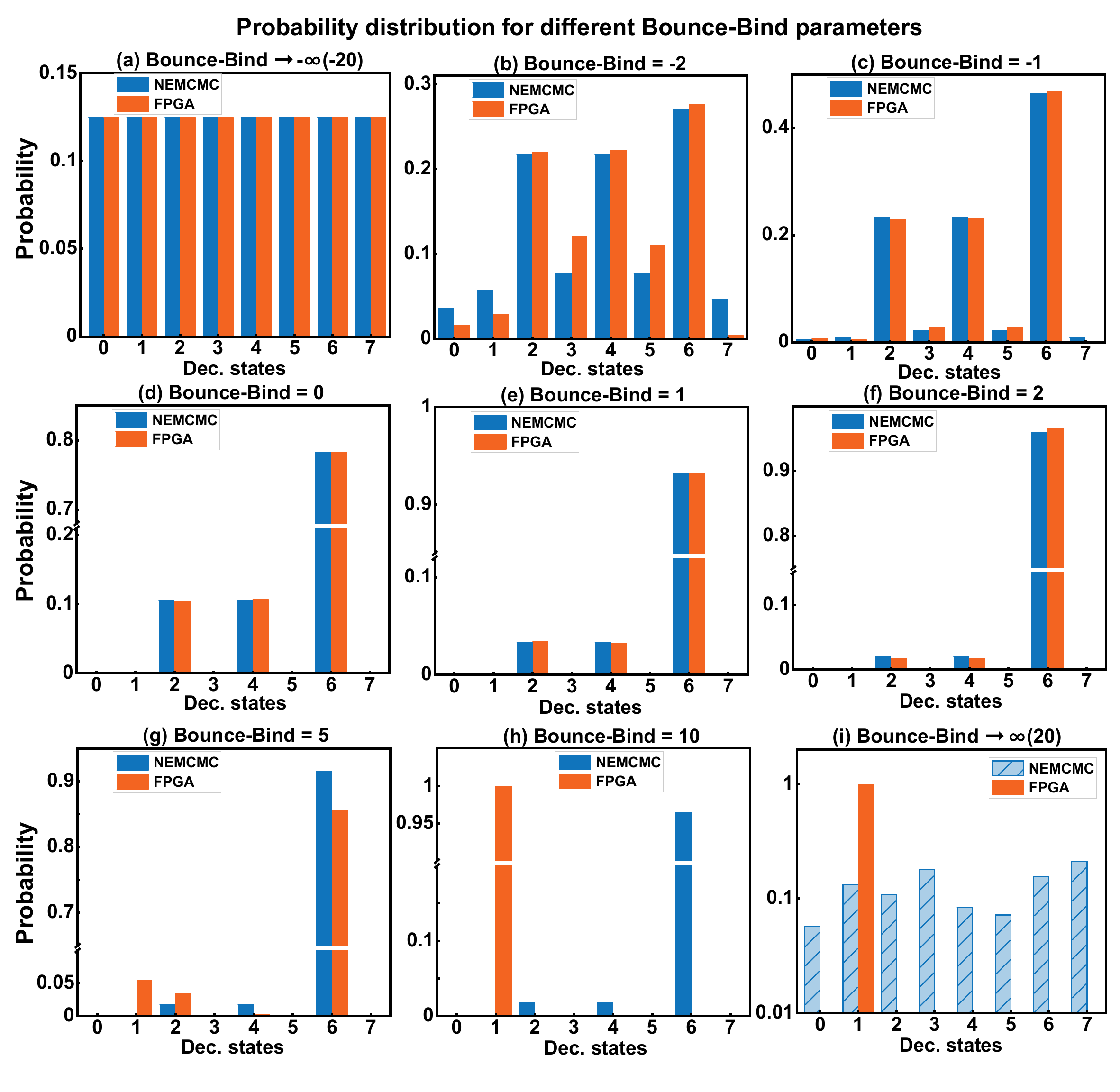}
\caption{\textbf{Probability distribution function under different Bounce-Bind parameters in the case of a 3-bit Ising problem.} We obtain the probability distribution obtained by FPGA with Gibbs sampling at $\beta = 1$ and $10^5$ sweeps under no annealing mechanism. The blue bars stand for the probability distribution calculated by Non-Equilibrium Markov Chain algorithm mentioned in Supplementary Section S1 and orange bars stand for the data obtained from FPGA. The blue shaded bars in \textbf{(i)} are only for illustration due to the failure of the theory of random process, whose probability distribution is related to the initial state of the system. }\label{pdf}
\end{figure}

Unlike the classic update process which is only related to the interaction matrix $\boldsymbol{J}$, the external field $\boldsymbol{h}$, and the state of the neighboring spins, the Bounce-Bind mechanism incorporates the current state $m_i$ and adjusts the state flipping frequency via parameter $\mathcal{B}$. This mechanism does not alter the energy landscape of the system. The ground state, excitation levels, and global minima remain fixed regardless of the value of $\mathcal{B}$. Instead, it directly controls how the system moves through its state space by modifying the local update rule. Without loss of generality, we assume $\mathcal{B}=\pm1$ to separately study the two mechanisms of our BBIM. When $\mathcal{B}=-1$,  the system exhibits bounce behavior, and the update process is:
\begin{equation}
    I_{Bounce,i} = I_i-m_i
\end{equation}
which implies that the next state has a tendency to flip relative to the current state. As shown in Fig. \ref{stp}(b) and comparing it with Fig. \ref{stp}(c), $\mathcal{B}=-1$ leads to a decrease in the probability of maintaining the current spin in the next state, thereby facilitating the state transition, which reflects the ``bounce" behavior of the Bounce-Bind term, akin to a tennis ball bouncing away from its current position. The system becomes restless: spins flip frequently, allowing it to escape shallow local traps. However, this same behavior prevents the system from settling into the lowest-energy state. Even though the ground state is still energetically favored, the constant flipping reduces the time it spends there, resulting in a lower probability of observing the system in its ground state, as seen in Fig. \ref{pdf}(c) compared to Fig. \ref{pdf}(d).

When $\mathcal{B} \to -\infty$, the update function simplifies to:
\begin{equation}
    m_i(t+1)=-m_i(t).
\end{equation}
 In this extreme case, every spin flips to its opposite value at each step. This will cause the system state to reverse based on the initial state, leading to a situation where it keeps flipping, preventing the system from converging. As shown in Fig. \ref{stp}(a), the transition matrix becomes a permutation matrix that swaps each state with its bitwise inverse. Despite this extreme restriction, Fig. \ref{pdf}(a) shows a uniform probability across all $2^N$ configurations. This uniformity arises not from exploration, but from averaging over many independent runs, each initialized from a different, randomly chosen starting state. Since every initial state is sampled with equal frequency, the resulting probability distribution is uniform across all configurations.

When $\mathcal{B} = 1$, it operates in the binding regime, and the state is updated by:
\begin{equation}
    I_{Bind,i}=I_i+m_i
\end{equation}
which indicates that the likelihood of maintaining the current state is enhanced as shown in Figs. \ref{stp}(c) and (d), that inhibits the flipping of the spin. On the other hand, this will facilitate the rapid convergence of the system. This characterizes the``Bind" behavior of the BB term, akin to a shot put sticking to its current state due to inertia.  The system rapidly loses mobility: once a spin aligns with its local field, it resists flipping, even if the global energy could be lowered. This leads to faster convergence toward stable configurations. And since the ground state typically has the strongest local bias, the probability of the system being found in it increases significantly, which is shown in Fig. \ref{pdf}(e). When $\mathcal{B} \to \infty$, the next state is obtained by:
\begin{equation}
    m_i(t+1)=m_i(t)
\end{equation}
which leads to the freezing of system's state which only depends on the initial state. As shown in Fig. \ref{stp}(f), the transition matrix becomes a perfect identity that every state maps only to itself. There is no dynamics left, only memory. The final state is determined entirely by the initial condition. Thus, it is impossible to calculate the distribution as the blue shaded bars in Fig. \ref{pdf}(i).  The system doesn’t evolve toward the ground state; it carries forward whatever state it began in. 

Thus, the Bounce-Bind term acts as a dynamical knob: At $\mathcal{B}=-1$, it makes the system overactive, flipping too often to settle. At $\mathcal{B}=1$, it makes the system overstable, freezing before it can find the best solution. In both extremes, the energy landscape remains unchanged, and only the motion through it is altered.

Moreover, we initialize the BBIM in the state 1 ($\braket{001}$) to examine how the system evolves toward its long-term behavior under different Bounce-Bind conditions, as shown in Figs. \ref{pdf}(f) to (i). For moderate values of $\mathcal{B}<2$, the probability distribution obtained through the Markov stochastic process shows a clear increase in the likelihood of the ground state. This reflects a regime in which the system balances exploration and stability, allowing it to escape shallow traps while gradually settling into low-energy configurations. However, when $\mathcal{B}$ exceeds 2, the probability of the ground state begins to decline. Instead, the system increasingly retains its initial configuration. By $\mathcal{B}=10$, as illustrated in Fig. \ref{pdf}(h), the system becomes effectively locked into the initial state $\braket{001}$, with negligible transitions to any other state. As $\mathcal{B}$ approaches infinity, the state transition matrix converges to the identity matrix, shown in Fig. \ref{stp}(f). In this limit, the stationary equation 
\begin{equation}
    \boldsymbol{\pi} \mathbb{P}=\boldsymbol{\pi}
\end{equation}
 holds for any probability distribution $\boldsymbol{\pi}$ where $\mathbb{P}$ is the transition probability in Fig. \ref{stp}, meaning the system has lost all memory of energy and is frozen in its initial condition. Consequently, the theoretical steady-state distribution becomes ill-defined, as depicted in Fig. \ref{pdf}(i).

To demonstrate these behaviors in a practical context, we apply the BBIM to an image search problem formulated as a MAX-CUT instance. The evolution of the solution is captured in an animation provided in the supplementary materials. When $\mathcal{B}$ is positive, the system converges rapidly to a stable image within a few iterations. However, the resulting image is often blurry or corrupted, indicating that the system has settled into a poor local minimum. This is consistent with the Bind mechanism, which suppresses flips even when they lead to lower energy. In contrast, when $\mathcal{B}$ is negative, the system explores more broadly across configurations. Convergence is slower, but the final image is sharper and more accurate, reflecting a higher-quality solution. When the magnitude of negative 
$\mathcal{B}$ becomes too large, for example $\mathcal{B}=-2$, the system enters a persistent oscillatory state, where spins flip back and forth without settling. The image flickers uncontrollably, consistent with the Bounce mechanism driving the system away from any stable configuration.

Note that the BBIM can be naturally extended to higher-order Ising machines by incorporating third-order interactions,
\begin{equation}
    I_{BB, i} = \sum_{j<k}J_{ijk}^{(3)}m_jm_k+\sum_jJ_{ij}^{(2)}m_j+h_i + \mathcal{B}m_i
\end{equation}
where $J_{ijk}^{(3)}$ models the third-order interaction and $J_{ij}^{(2)}$ models the second-order interaction.  The term $\mathcal{B}m_i$ acts as a dynamic bias, independent of the energy landscape, to modulate the tendency of each spin to persist or flip.
  
In summary, the Bounce-Bind term controls the search trajectory of the Ising machine by modifying the local update rule. When $\mathcal{B}$ is positive, the spin is encouraged to maintain its current state, leading to rapid convergence but an elevated risk of becoming trapped in suboptimal configurations. When $\mathcal{B}$ is negative, the spin is biased to reverse its current state, promoting broader exploration at the cost of slower convergence. This mechanism does not rely on thermal noise or energy differences. It is a direct modulation of state persistence, a form of dynamical engineering that operates outside the constraints of equilibrium statistical mechanics.

\subsection{FPGA implementation}

\begin{figure}[h]
\centering
\includegraphics[width=1\textwidth]{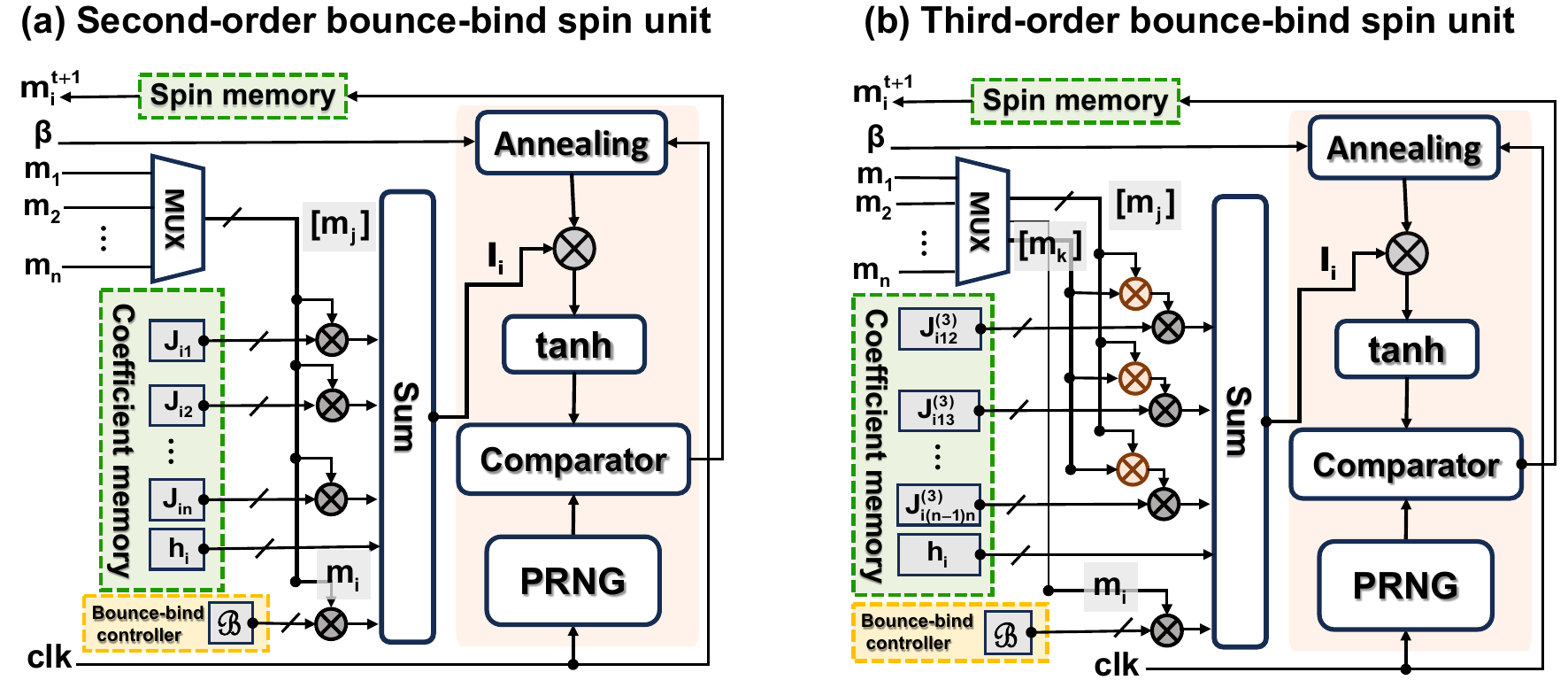}
\caption{\textbf{The basic spin architecture of Bounce-Bind Ising machine implemented by FPGA.} We have respectively designed Ising machines based on the second-order Ising model \textbf{(a)} and third-order Ising model \textbf{(b)}. $J$ and $h$ are only represented by 2 bits. In principle, they can also be represented by only one bit ($0$ for $-1$ and 1 for $+1$). When applied to the XORSAT problem, the bit width of $J$ and $h$ is 3 based on the Method \ref{mtd1}. For the third-order Ising machine, we mainly apply it to the 3R3X problem. In fact, it does not involve the second-order coupling term, so only the third-order coupling term is retained, and only two bits are used to represent $J$ and $h$. In \textbf{(b)}, the brown multiplier represents the multiplication of two spins in the same clause. To reduce hardware cost, we used an XOR gate to implement this logic instead of using an $1\times1$ multiplier. Moreover, when $\mathcal{B}=0$, the Bounce-Bind spin unit reduces to the classical one.}\label{spins}
\end{figure}

Our experiments were conducted using the Xilinx Vivado suite on the Xilinx Virtex-7 XC7VX485TFFG1157-1 FPGA. For the Bounce-Bind parameter $\mathcal{B}$, we adopted a signed fixed-point precision format \textbf{s[2][3]}, which uses 1 bit for the sign, 2 bits for the integer part and 3 bits for the fractional part. Moreover, we use a 32-bit \textbf{Linear} \textbf{Feedback} \textbf{Shift} \textbf{Register} (LFSR) with an XNOR feedback to the first register (0) using taps from registers 32, 22, 2 and 1 \cite{peter1998efficient} to generate random numbers. The architecture of spin in our Ising machine is illustrated in Fig. \ref{spins}. The k-nearest neighbor spin interaction structure of our proposed Ising machine is equivalent to the ($k+1$)-nearest neighbor structure of the traditional CMOS Ising machine in terms of effective interaction range. Specifically, for dense graphs, the two Ising machines exhibit nearly identical resource consumption.

Although the Bounce-Bind term does not alter the energy landscape or the identity of the ground state, it fundamentally reshapes the dynamical pathway by which the system evolves. According to the Boltzmann distribution, the stationary probability of the ground state should remain unchanged if the system thermalizes. However, the Bounce-Bind mechanism operates outside the framework of detailed balance, and therefore the system may not reach the theoretical Boltzmann distribution. Instead, the final state probability distribution reflects a non-equilibrium steady state determined by the interplay of energy gradients and the bias $\mathcal{B}$. Although after the Bounce-Bind mechanism is introduced, the sampling process designed for the Ising model based on the Boltzmann distribution is no longer balanced (we call this a non-equilibrium Markov chain), for the FPGA implementation based on the Ising machine, we still use Glauber dynamics or Gibbs sampling to study this process, and through experiments and data, we prove that this process is still applicable to our BBIM.

To quantify this effect, we compared the output of our FPGA-based BBIM with a numerical simulation based on Non-Equilibrium Markov chain as detailed in Supplementary Section S1. For a small-scale 3-spin problem corresponding to Fig. \ref{stp}, we computed the probability distribution of all possible states using both methods. When $\mathcal{B}$  is within a moderate range, the results from our hardware implementation and the theoretical stochastic process agree within acceptable statistical error bounds. However, for large $\mathcal{B}$, a clear divergence emerges: the hardware system becomes increasingly trapped in its initial state, while the Gibbs sampler continues to sample according to the energy landscape. This discrepancy confirms that the Bounce-Bind term alters the convergence dynamics without altering the underlying energy function.

In addition, it should be noted that we referred to the design in Ref.  \cite{patel2020ising} and introduced the ``hitting engine" to obtain the results of each run of the Ising machine. We employed the ``hitting time"\cite{patel2022logically} criterion to terminate the sampling process instead of ``mixing time"\cite{bremaud2001markov,levin2017markov}, that is, to determine the state of the solution based on the Ising energy (the lowest energy represents the ground state). 

\subsection{Benchmarks}
To ensure completeness and validation of the experimental data, we chose dense MAX-CUT problems and sparse 3R3X problems as well-established benchmarks. The Ising models for these two types of problems are detailed in Method \ref{mtd1}. We benchmark instances of 3R3X problems (degree = 9) with node counts increasing from 16 to 160 in increments of 16 for second-order SAT. In addition, we verified our theory and model on the third-order SAT problems. We benchmark instances of dense MAX-CUT problems (edge density = 0.5) with node counts increasing from 10 to 200 in increments of 10. For 3R3X problems, we utilized the second-order SAT dataset from Ref. \cite{nikhar2024all} and employed the details from Method \ref{mtd2} to generate the third-order dataset. For each size of the MAX-CUT problem, we randomly generated 100 instances that belong to Erd\H{o}s--R\'enyi  $G(N, \frac{1}{2})$ mentioned in Ref. \cite{hamerly2019experimental}, and each instance was executed 100 times to ensure statistical significance. Problem instances of size $\leqslant$ 150 were solved by BiqCrunch 2.0\cite{krislock2017biqcrunch} and validated by the exact solver BiqMac\cite{wiegele2007biq} to serve as the ``ground truth solution".  For instances of size $>$ 150, which could not be solved within reasonable time limits, we ran the machine sufficiently long and regarded the lowest energy found as the approximate ground state. In addition, we conducted verification on the larger-scale MAX-CUT graphs, which are derived from G-set (including G22 and G39) \cite{helmberg2000spectral} and K2000 generated by the machine-independent graph generator\cite{rudy}.

For small-scale MAX-CUT instances and satisfiability (SAT) problems, the ground state energies are known. In these cases, the system is considered to have found a valid solution when its computed Ising energy matches the known ground state energy. Upon reaching this condition, the simulation or hardware run terminates early, thereby improving computational efficiency without compromising solution accuracy.

For large-scale MAX-CUT problems, the exact ground state energy is typically unknown due to the exponential growth of the state space. In such scenarios, we record the highest cut value observed during a fixed number of iterations, using the hitting time criterion as defined in our implementation: the first occurrence at which the lowest energy state is attained.

Moreover, in the following discussion, we also use \textbf{sampling rounds} interchangeably to denote the number of samples. For a problem with \( N \) nodes, when the number of sampling rounds is \( \textbf{Rounds} \), the total number of samples is given by \( N_{\text{samples}} = N \times \text{Rounds} \). This formula will also be used in subsequent metric calculations.

\subsection{Effect of Bounce-Bind parameter on success probability}

\begin{figure}
\centering
\includegraphics[scale=0.4]{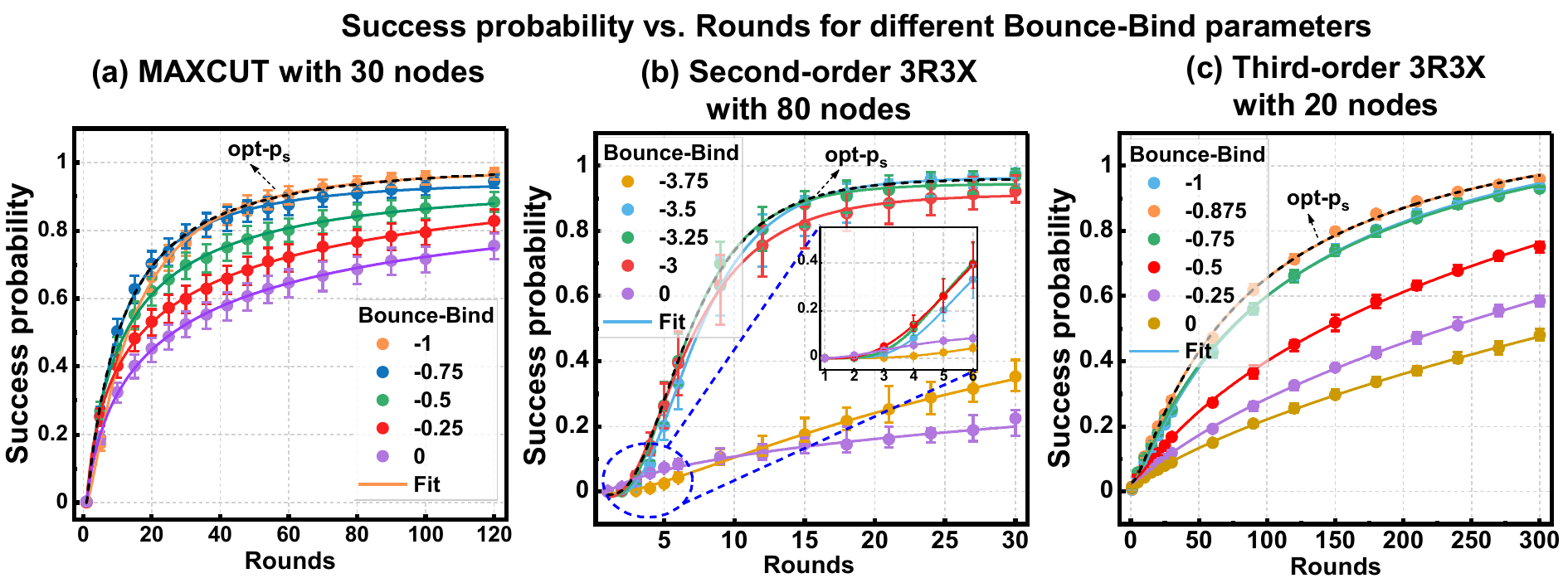}
\caption{\textbf{Success probability versus rounds taken for different Bounce-Bind parameters.} We provide three examples to illustrate how the success probability varies with the number of samples under different Bounce-Bind parameters, where the black dashed line represents the maximum success probability at given sampling rounds. For optimal success probability, $\mathcal{B}$ needs to be carefully tuned for every size N and benchmark for the chosen sweeps. Data points are averaged over 100 random instances and 1 000 trials per instance, each trial taking the sweeps mentioned in the figure. Error bars on each of the graphs were generated by calculating the bootstrap resamples, with $95\%$ confidence interval for the parameter being estimated. In \textbf{(b)}, we zoom in on the figure ranging from 0 to 5.
} 
\label{sp_for_bb}
\end{figure}

 The choice of $\mathcal{B}$ has a strong effect on algorithm performance and should be characterized to select optimal parameters for different problem instances. Our focus is not on discussing the parameters of existing solvers including $\beta$ and so on, so we adopt a fixed annealing schedule without fine-tuning according to the specific size and characteristics of the problem\cite{isakov2015optimised}. In our annealing algorithm, the initial inverse temperature $\beta_0$ is set to 0.125, using a linear annealing schedule with an increment of 0.125, which is suitable for our FPGA implementation, and the search stops when $\beta = 4$.

\begin{figure}
\centering
\includegraphics[scale=0.4]{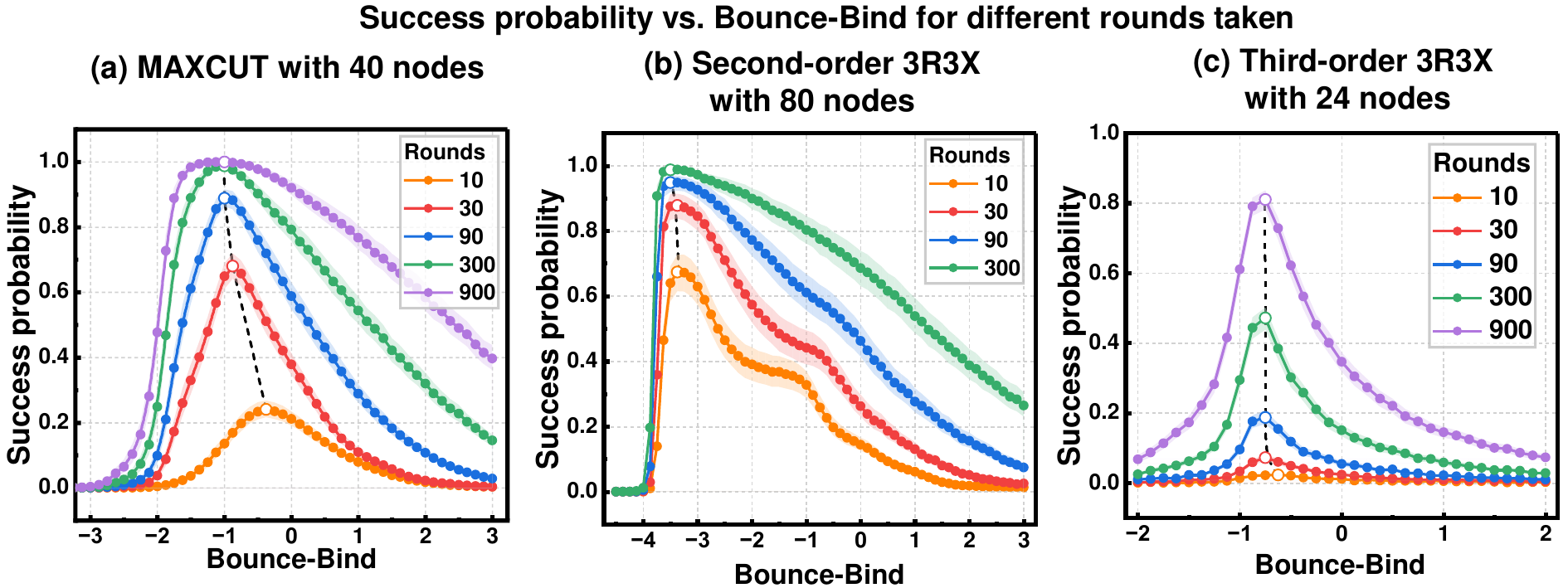}
\caption{\textbf{Success probability versus Bounce-Bind parameter taken for different search rounds.} For these three types of problems, we used one example each to illustrate how the success probability varies with the Bounce-Bind parameters across different sampling rounds. Data points are averaged over 100 random instances and 1 000 trials per instance, each trial taking the MCSs mentioned in the figure. Error bars on each of the graphs were generated by calculating the bootstrap resamples, with $95\%$ confidence interval for the parameter being estimated.
} 
\label{sp_for_smps}
\end{figure}

\begin{figure}
\centering
\includegraphics[scale=0.4]{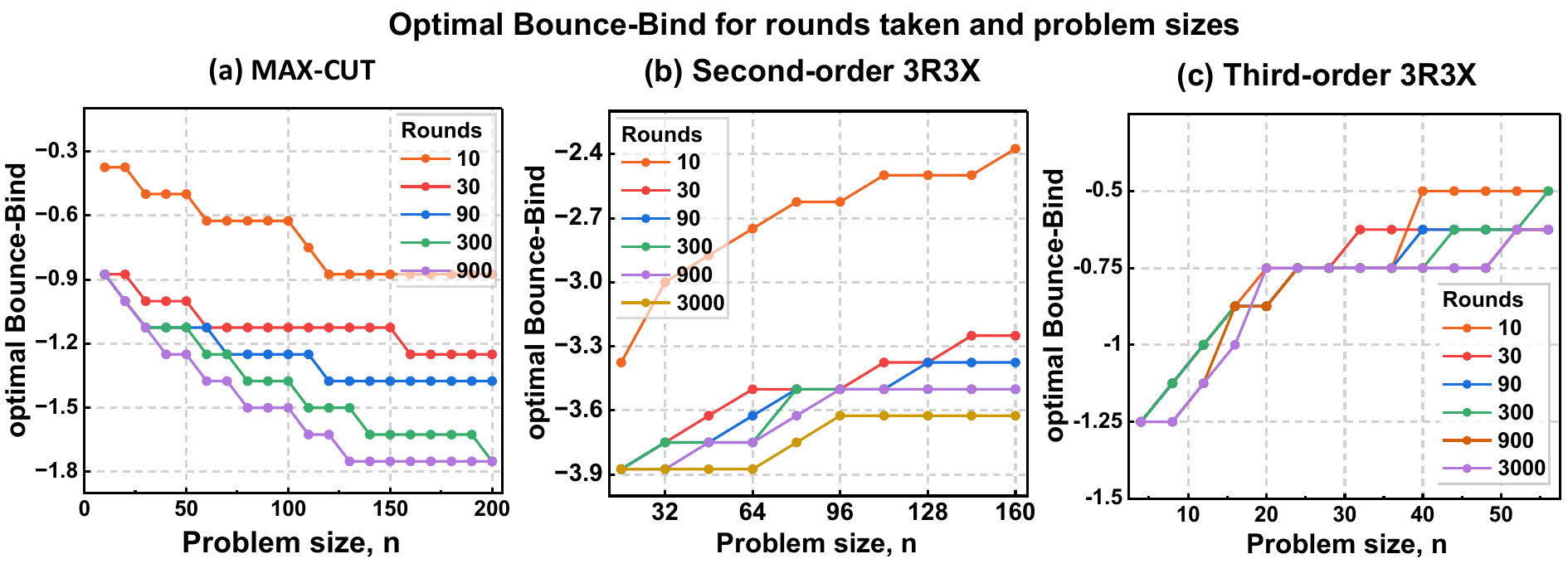}
\caption{\textbf{Optimization of Bounce-Bind parameters.} We provided reference values for the optimal Bounce-Bind parameter for three different problem instances at varying values of \( N\). Since our scan over $\mathcal{B}$ values was in increments of 0.125, the obtained optimal $\mathcal{B}$ values are not precise but instead fluctuate around the theoretical optimum. For the MAX-CUT problem, the trend of the optimal $\mathcal{B}$ value with respect to problem size N differs from that of the other two problems, as a consequence of SAT's sparse connectivity versus MAX-CUT's dense structure.
} 
\label{opt_bb}
\end{figure}

According to Subsection \ref{bbeft}, a negative $\mathcal{B}$ is conducive to increasing the success probability of the BBIM, which is verified in Fig. \ref{sp_for_bb} and Fig. \ref{sp_for_smps}. When $\mathcal{B}$ is an appropriate negative value (which means $\mathcal{B}$ has an optimal value), the Bounce mechanism greatly increases the success probability of the Ising machine. Meanwhile, the optimal value of $\mathcal{B}$ is a function of the samples. As shown in Fig. \ref{sp_for_smps},  given a fixed round, we can observe clear peaks in the probability of finding the ground state. Moreover, for a fixed size problem, the peak position of $\mathcal{B}$ gradually shifts leftward as the number of sampling rounds increases, which means that the optimal value of $\mathcal{B}$ progressively decreases. For second-order and third-order SAT, as the size of the problem increases, the position of this peak gradually shifts to the right, which means that the optimal value of $\mathcal{B}$ progressively increases in Figs. \ref{opt_bb}(b) and (c). In contrast, for MAX-CUT, the trend of the optimal value of $\mathcal{B}$ with respect to problem size is opposite, showing a gradual decrease as the problem size increases in Fig. \ref{opt_bb}(a). This is because, in the benchmark problems we selected, SAT corresponds to a sparse graph while MAX-CUT corresponds to a dense graph. Specifically, the maximum degree in SAT is 9, whereas for MAX-CUT with an edge density of 0.5, the degree scales proportionally with the problem size. When the number of sampling rounds is fixed, the optimal value of $\mathcal{B}$ depends not only on the problem size but also on the degree of the Ising graph. Based on the above results, the optimal value of $\mathcal{B}$ increases with larger problem sizes but decreases with higher graph degrees. For SAT instances, problem size is the dominant factor, whereas for MAX-CUT instances, the graph degree plays the dominant role. Overall, the performance across benchmark problems exhibits a clear dependence on the bias parameter $\mathcal{B}$: it remains low below a critical threshold, rises to a maximum at an optimal value, and gradually declines for larger values.


Moreover, the optimal $\mathcal{B}$ changes with the number of samples taken, as shown in Fig. \ref{sp_for_smps}. This is because when the number of samples is small, insufficient annealing time on the Ising machine causes the bounce mechanism to have a negative effect: namely, the system state keeps changing, thus prolonging the path to the ground state, which is particularly evident in Fig. \ref{sp_for_smps} (a). Meanwhile, the Bind mechanism promotes rapid convergence of the system, preventing it from traversing a sufficient number of states to reach the ground state. As a result, performance is poorer with increasing $\mathcal{B}$ when $\mathcal{B}$ is positive.

\subsection{Effect of Bounce-Bind parameter on time to solution}
Since the search process of the Ising machine belongs to a heuristic algorithm, the probability of finding the ground state can be increased by increasing the number of sweeps, increasing the number of repetitions, or using both in parallel. To achieve this goal and make the optimal choice with the minimum total computational load, we need to balance the relationship between the number of sweeps (search duration) and the number of repetitions (search times). We use the Time-to-Solution (TTS) metric to quantify the variation pattern of the computing time of the Ising machine with the problem scale when seeking the optimal solution. If the probability of obtaining the ground state in a single repetition experiment when the running time is $t_f$ is $p_s(t_f)$, the total number of repetitions R required to find the ground state with 99\% and TTS is defined as:
\begin{equation}
    \text{TTS} = t_fR=t_f\frac{\log(1-0.99)}{\log(1-p_s(t_f))}
\end{equation}
In our implementation, $t_f$ is given by $t_f=N_{samples}\cdot N_{clks}/f_{clk}$, where $N_{clks}$ is the number of clock cycles required for each sampling and $f_{clk}$ is the clock frequency (210 Mhz for our FPGA implementation). For the 3R3X problem, we use the optimal TTS (optTTS) for a quantile q to evaluate the solver's computational time:
\begin{equation}
    \langle \text{TTS} \rangle_q = \min_{t_f} \langle t_fR \rangle_q
\end{equation}
We use Bayesian inference to estimate $p_s$, and then use bootstrap resample to determine the average of a quantile $q$ of the TTS, established by Kowalsky et al.\cite{kowalsky20223}. For 3R3X problem, the general scaling behaviour is observed to follow the model:
\begin{equation}
    \langle \text{TTS} \rangle_q \sim 10^{\gamma n+\eta}
\end{equation}
$\gamma$ and $\eta$ are the fitting parameters, and lower values of $\gamma$ and $\eta$ indicates that the Ising machine has better performance.

\begin{figure}
\centering
\includegraphics[scale=0.28]{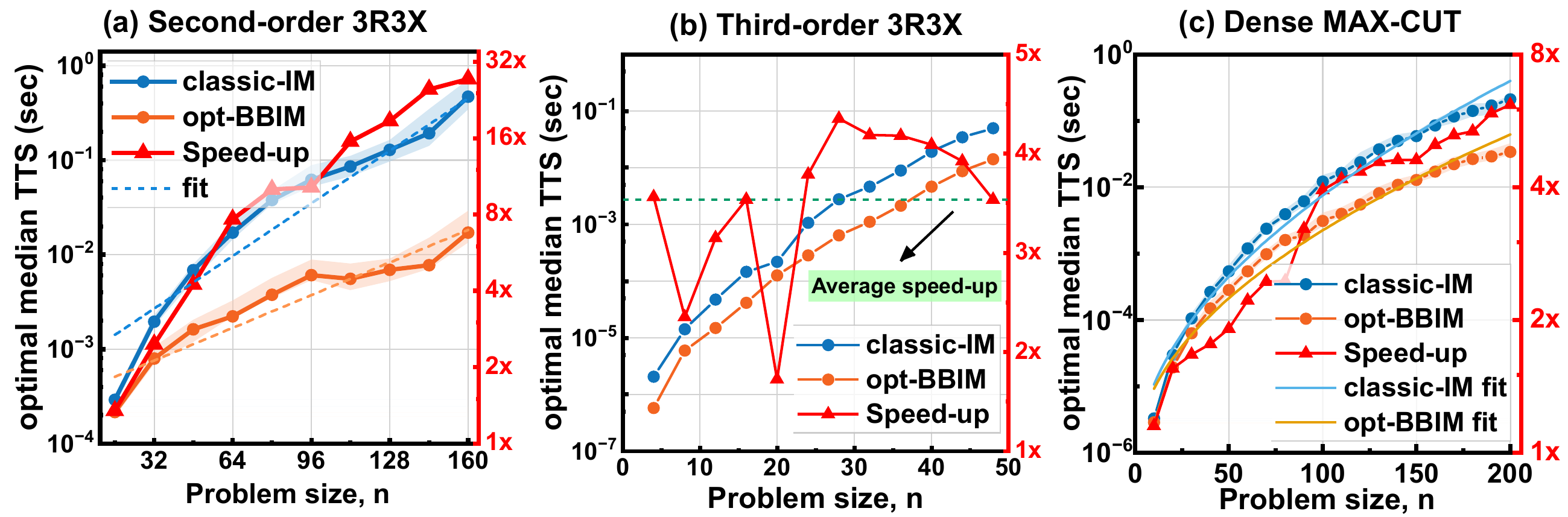}
\caption{\textbf{Comparison of optimal median time to solution.} In this figure, we respectively use the classical Ising machine and BBIM to solve the second-order, third-order 3R3X problems , and dense MAX-CUT problems and compare their optimal median TTS, as well as showing the acceleration ratio. Blue dots represent the measured data for the Ising machine based on classic Ising model, orange dots represent the measured data for the Ising machine based on our Bounce-Bind Ising model and red triangles represent the speed-up of these two. Error bars are obtained from 95\% confidence intervals by bootstrap resamples. Each data point contains 100 instances, and each instance run 1,000 times.} 
\label{3r3x_tts}
\end{figure}

We compared the time to solution of FPGA-based Ising machines based on the classic Ising model (classical-IM) and the optimal Bounce-Bind Ising model (opt-BBIM). The performance parameters of the optimal Bounce-Bind Ising machine are derived from the metrics when the Bounce-Bind parameter is optimized. The TTS fitting behaviors of the two Ising machines are shown in Fig. \ref{3r3x_tts} and the fitting parameters for 3R3X are shown in Table \ref{tab_sat}. Since our Ising machines adopt the sequential update strategy, the performance of classical-IM is not as good as the highly parallelized search p-computer based on graph coloring in Ref. \cite{nikhar2024all}. However, the opt-BBIM achieves a greater acceleration compared to classical-IM under almost the same hardware resource overhead. Our opt-BBIM even has improved performance compared to p-computer, which is shown in Table \ref{tab_sat}. However, the opt-BBIM cannot fit well using the exponential model $10^{\gamma n+\eta}$ for 3R3X, while the fitting behavior using the exponential square root model $e^{\alpha \sqrt{n}+\beta}$ or the power function model $an^k$ is better. This is because as the size of problem increases, the optimal-$\mathcal{B}$ when obtaining the optimal TTS is not fixed, as shown in Fig. \ref{sp_for_smps}(a). This suggests that it is precisely this correlation between the Bounce-Bind parameter and the problem size n that causes this non-exponential correlation. Despite this, compared with the classical-IM, the BBIM still shows a significant solution acceleration, and when solving the second-order 3R3X problem, it achieves an acceleration ratio of 1.35 to 27.3 times ($27.3\times$ is taken when n=160), and this speedup ratio is expected to continue to increase with problem scale.

\begin{table}[h]
\caption{Comparison of the fitting parameters $\gamma$ and $\eta$ of different solvers}\label{tab_sat}
\begin{tabular*}{\textwidth}{@{\extracolsep\fill}lcccccc}
\toprule%
& \multicolumn{3}{@{}c@{}}{$\gamma$} & \multicolumn{3}{@{}c@{}}{$\eta$} \\\cmidrule{2-4}\cmidrule{5-7}%
Solver & $q=0.25$ & $q=0.5$ & $q=0.75$ & $q=0.25$ & $q=0.5$ & $q=0.75$ \\
\midrule
SATonGPU  & n/a & 0.0171(7) & n/a & n/a & -5.9(3) & n/a \\
\midrule
DAU   & 0.0181(2) & 0.0185(4)  & 0.0190(4) & -3.51(4) & -3.56(7) & -3.49(7)\\
\midrule
SBM   & 0.0211(7) & 0.0217(6)  & 0.0234(8) & -2.6(2) & -2.6(1) & -2.7(1)\\
\midrule
PT    & 0.0239(1) & 0.0248(2)  & 0.0252(1) & -0.92(2) & -0.97(4) & -0.97(2)\\
\midrule
MEM   & 0.030(9) & 0.025(2)  & 0.024(3)  & -1(2) & -0.6(2) & -0.2(2)\\
\midrule
DWA   & n/a & 0.08(4)  & n/a  & n/a & -6(2) & n/a\\
\midrule
p-bits (FPGA) & 0.0194(1) & 0.0206(2) & 0.0210(3) & -3.70(4) & -3.76(6) & -3.74(2) \\
\midrule
\textbf{classical-IM} & \textbf{0.0207(0)} & \textbf{0.0224(2)} & \textbf{0.0249(7)} & \textbf{-3.12(1)} & \textbf{-3.02(3)} & \textbf{-2.98(5)} \\
\midrule
\textbf{(FPGA)} &  &  &  &  &  &  \\
\midrule
\textbf{opt-BBIM} & \textbf{0.0187(6)} & \textbf{0.0189(4)} & \textbf{0.0187(4)} & \textbf{-3.67(5)} & \textbf{-3.43(3)} & \textbf{-3.12(6)} \\
\midrule
\textbf{(FPGA)} &  &  &  &  &  &  \\
\botrule
\end{tabular*}
\footnotetext{The data is derived from Table 2 of Ref. \cite{kowalsky20223} and Table 1 of Ref. \cite{nikhar2024all}, and the data highlighted in bold is obtained by the BBIM and classic Ising Machine implemented by FPGA. We implement the Ising machine based on the classic Ising model and the Bounce-Bind Ising model, and the BBIM has lower $\gamma$ and $\eta$, which represents better performance.}
\end{table}

However, the results for the third-order 3R3X problem in Fig. \ref{3r3x_tts}(b) did not achieve an excellent performance improvement similar to that of the second-order problem, but still obtained a notable acceleration compared to the classical-IM, demonstrating the applicability of the Bounce-Bind mechanism to the third-order Ising problem. Importantly,  for the third-order 3R3X problem, both the classical-IM and opt-BBIM exhibit exponential scaling consistent with $10^{\gamma n+\eta}$.

The above results are based on the sparse Ising problem. Next, taking the MAX-CUT problem as the benchmark, we will conduct further verification on the dense Ising problem and attempt to apply it to larger-scale problems.
In Fig. \ref{3r3x_tts}(c), we show the results of benchmarking performance on the MAX-CUT whose edge density is 0.5. After introducing the Bounce-Bind mechanism, there is still the scaling of $O(e^{\sqrt{n}})$ for the MAX-CUT problem. Moreover, compared with the classical-IM, BBIM has a faster solution speed, and as the problem size increases, the speed-up ratio of the solution also increases accordingly (from $1.15\times$ at $n=10$ to $6.15\times$ at $n=200$).

\subsection{Performance on MAX-CUT with 2000 nodes}
\begin{figure} 
\centering
\includegraphics[scale=0.32]{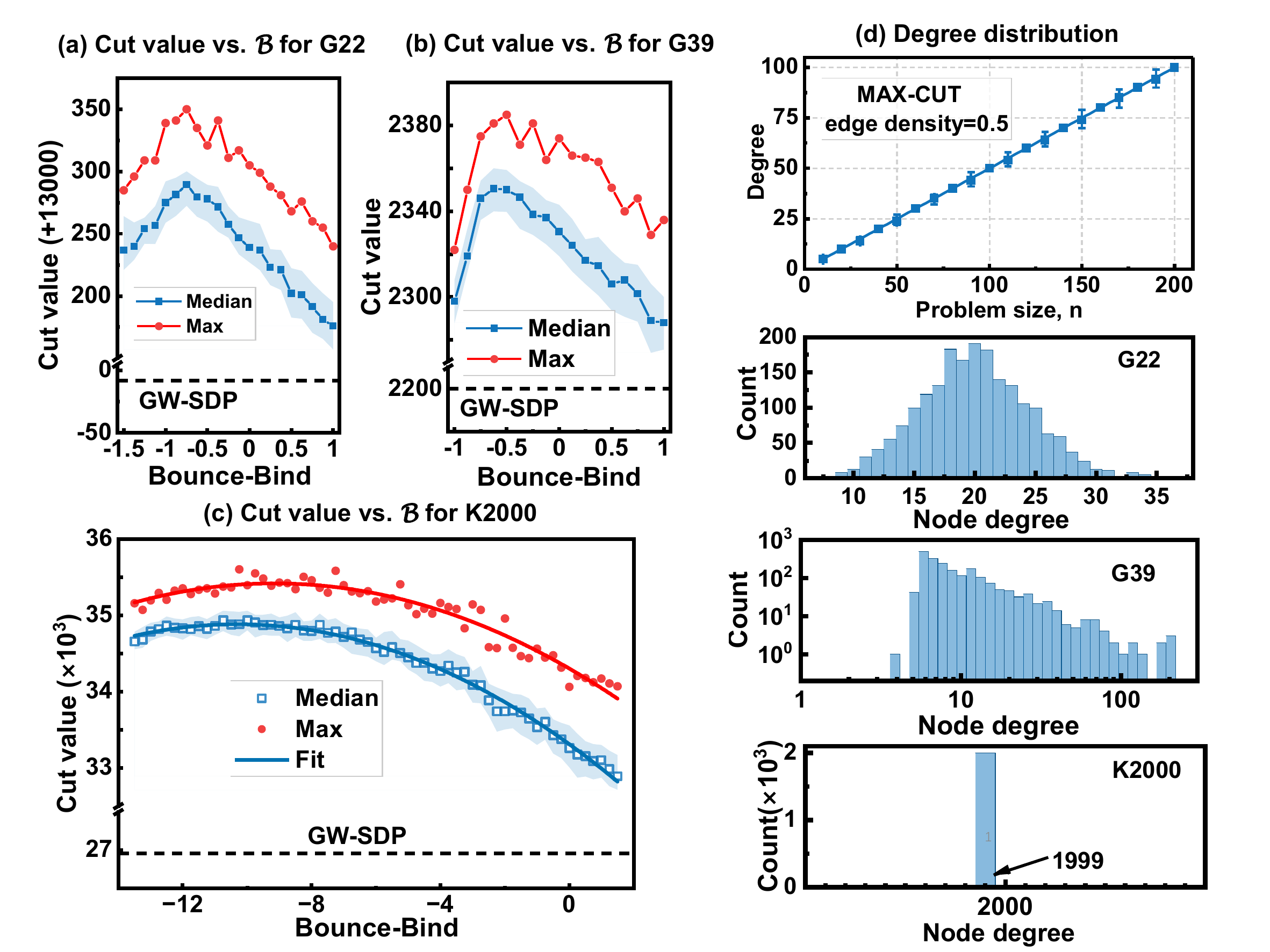}
\caption{\textbf{Performance and speed-up of TTS for MAX-CUT.} \textbf{(a)} to \textbf{(c)} Optimization of the Bounce-Bind parameter $\mathcal{B}$ on the MAX-CUT problem with 2000 nodes in 5 ms. Each data point represents the median or maximum value of 100 run results and shading indicates interquartile range (IQR; 25th, 75th percentile range of instances). \textbf{(d)} Degree distributions of G22, G39, and K2000, showing that G22 and G39 are sparse, while K2000 is dense.}
\label{maxcut_tts}
\end{figure}

 We also studied the approximate accuracy of the BBIM on 2000-node graphs from Gset (G22 and G39), and K2000 generated with the machine-independent graph generator in Ref. \cite{rudy}. As shown in Fig. \ref{maxcut_tts}(d), the K2000 is a complete graph with 1,999,000 undirected edges whose weights are randomly chosen from $\pm1$ while G22 and G39 are sparse graphs. The edge weights of G22 with 19,990 edges are all 1 and those of G39 with 11,778 edges, with weights from $\{-1, 1 \}$. In Figs. \ref{maxcut_tts}(a) to (c), we fix $t_f=5 ms$ for the BBIM and show the cut value for varying Bounce-Bind parameters. Obviously, the Bounce-Bind mechanism is also applicable to the MAX-CUT problem with 2000 nodes, and it is further verified that the optimal value of $\mathcal{B}$ decreases as the density of the graph increases. According to Figs. \ref{maxcut_tts}(a) to (c), we can obtain the empirical parameters of the optimal values of $\mathcal{B}$ for these three instances and apply them to Table \ref{tab_max}. In Table \ref{tab_max}, we compared several different solvers and solving algorithms (GW-SDP). We compute a scalable approximation of the GW-SDP by solving a low-rank semidefinite relaxation using the Burer–Monteiro formulation, followed by standard Goemans–Williamson randomized hyperplane rounding. The resulting cut value is used as a benchmark reference for K2000 while the cut values by GW-SDP for G22 and G39 are derived from Fig. 3 of Ref. \cite{inagaki2016coherent}. For these three instances, all solvers have achieved larger cut values compared to GW-SDP, and the BBIM outperforms SA and CIM in terms of cut values.

\begin{table}[h]
\caption{Comparison of cut values of MAX-CUT with 2000 nodes}
\label{tab_max}
\begin{tabular*}{\textwidth}{@{\extracolsep\fill}lccc}
\toprule
\textbf{Graph} & \textbf{G22} & \textbf{G39} & \textbf{K2000} \\
\midrule
Best Known      & \textbf{13359} & \textbf{2408} & --\\
\midrule
SA (Best)       & 13336(50ms) & 2384(50ms) & 34802(50ms)\\
\midrule
SA (Mean)       & 13298(50ms) & 2359(50ms) & 33985(50ms)\\
\midrule
CIM (Best)       & 13313(5ms) & 2361(5ms) & --\\
\midrule
CIM (Mean)       & 13248(5ms) & 2328(5ms) & --\\
\midrule
\textbf{BBIM (Best)} & \textbf{13359}(10ms) & \textbf{2403}(10ms) & \textbf{35732}(50ms) \\
\midrule
\textbf{BBIM (Mean)} & \textbf{13338}(10ms) & \textbf{2381}(10ms) & \textbf{35239}(50ms)\\
\midrule
GW-SDP                & 12992 & 2200 & 26957\\
\bottomrule
\end{tabular*}
\footnotetext{Cut values obtained for the MAX-CUT problem on 2000-node graphs with the CIM, SA, GW-SDP and our BBIM. The SA and CIM data for G22 and G39 are derived from Fig. 3 of Ref. \cite{inagaki2016coherent}. For K2000, the data comes from the SA algorithm implemented by ourselves and the BBIM. We fix $\mathcal{B}=-0.5$ for G22, $\mathcal{B}=-1$ for G39 and $\mathcal{B}=-8$ for K2000 in the BBIM. A larger cut value represents better performance. The running duration is shown in parentheses ($\cdot$).}
\end{table}

\section{Conclusion}\label{sec13}
In this paper, we introduce the Bounce-Bind mechanism, a novel dynamical control principle for Ising machines to balance exploration and exploitation to enhance both the convergence speed and solution quality. At small problem sizes, the observed performance gains are modest, consistent with a relatively smooth energy landscape, allowing the system to approach optimal configurations with limited exploration. As the problem scale increases, the Bounce component becomes increasingly influential, facilitating broader exploration and improving solution quality, consistent with a growing ruggedness in the energy landscape. We have validated this mechanism on MAX-CUT problems, second-order 3R3X and third-order 3R3X problems, and large-scale MAX-CUT instances with up to 2000 nodes, consistently demonstrating that the Bounce mechanism significantly enhances both solution quality and solving speed. Compared to classical-IM, our BBIM achieves a speedup ranging from $1.35\times$ to $27.3\times$  on second-order 3R3X problems, an average  $3.5\times$ speedup on third-order 3R3X problems, and a speedup between $1.15\times$ and $6.15\times$ on dense MAX-CUT instances. Moreover, on the 2000-node MAX-CUT problem, BBIM achieves significantly higher cut values than SA, CIM, and GW-SDP. The consistent improvement across scales suggests BBIM is well-suited for large-scale optimization, with potential for further gains as problem size increases.

Although the optimal value of $\mathcal{B}$ depends on factors such as problem size $N$, graph density, coupling strength $J$, and interconnection topology (analogous to copy edge optimization in sparsification algorithms\cite{sajeeb2025scalable, patel2020ising}), Fig. \ref{opt_bb} shows that a significant performance improvement can be achieved with $\mathcal{B}=-0.5$ or $-1$. Thus, for most problems, these empirical values can be adopted as default settings.

We present a mechanism that can be integrated into Ising machines, rather than being confined to a specific algorithm. This mechanism can also be incorporated into other algorithms or solvers for COPs, further accelerating solution speed and improving solution quality.

\section{Methods}\label{method}
\subsection{The second-order and third-order Ising model}\label{mtd1}
The Ising machine, which is based on the Ising model, utilizes a physical arrangement of n spins to carry out a ground state search. Each spin in the machine adopts two states: up ($+1$) and down ($-1$), influenced by spin interactions (governed by matrix $\boldsymbol{J}$) and an external magnetic field ($\boldsymbol{h}$). For a spin configuration $\boldsymbol{m}=\{m_1, m_2, \cdots, m_n\}$, the Ising energy is expressed as:
\begin{equation}
E(\boldsymbol{m})=-\sum_{i<j}{J_{ij}m_im_j-\sum_{i}{h_im_i}} \label{method_label_1}
\end{equation}
And the third-order Ising model is:
\begin{equation}
    E(m) = -\sum_{i<j<k}J_{ijk}^{(3)}m_im_jm_k-\sum_{i<j}J_{ij}^{(2)}m_im_j-\sum_ih_im_i\label{method_label_2}
\end{equation}
In our Bounce-Bind model, we introduce a Bounce-Bind term to Eqs. (\ref{method_label_1}) and (\ref{method_label_2}):
\begin{equation}
    E(m) = -\frac{\mathcal{B}}{2}\sum_im_i^2 - \sum_{i<j}J_{ij}m_im_j - \sum_ih_im_i
    \label{method_label_3}
\end{equation}
\begin{equation}
    E(m) = -\sum_{i<j<k}J_{ijk}^{(3)}m_im_jm_k-\frac{\mathcal{B}}{2}\sum_im_i^2-\sum_{i<j}J_{ij}^{(2)}m_im_j-\sum_i h_im_i
    \label{method_label_4}
\end{equation}

\subsection{Ising model of benchmark problems}\label{mtd2}
We select the MAX-CUT problem with an edge density of 0.5 as the benchmark problem for the dense second-order Ising model. Given the weights of the edges $w_{ij}$, we can get the Ising model corresponding to the MAX-CUT problem:
\begin{equation}
    E(\boldsymbol{m}) = -\sum_{i>j}J_{ij}m_im_j=\sum_{i>j}w_{ij}m_im_j
\end{equation}
The cut value is given by:
\begin{equation}
    Cut(\boldsymbol{m}) = -\frac{E(\boldsymbol{m})}{2}-\frac{1}{4}\sum_i^n\sum_j^nJ_{ij}=-\frac{E(\boldsymbol{m})}{2}+\frac{1}{4}\sum_i^n\sum_j^nw_{ij}
\end{equation}

Moreover, we can easily obtain the ground-state energy of a problem of n variables 3R3X-SAT. The gadget allows us to cast the n-bit 3R3X problem as two-body Ising models with 2n spins, as each clause adds one auxiliary spin to the problem, and an n-bit instance has n clauses:
\begin{equation}
\begin{aligned}
G_{i}=& -h_s(m_{i1}+m_{i2}+m_{i3})+h_am_{ia} \\ 
&-J_s(m_{i1}m_{i2}+m_{i1}m_{i3}+m_{i2}m_{i3})-J_a(m_{i1}+m_{i2}+m_{i3})m_{ia} 
\end{aligned}
\end{equation}
where ($h_s, h_a, J_s, J_a$) can be ($-1, -2, 1, 2$) or ($-1, 2,
1, -2$), giving in both cases a minimal cost of -4 (there are other gadgets that can be used equivalently). In addition, we also consider the third-order Ising model for the 3R3X-SAT problem, which is more concise compared to its second-order counterpart:
\begin{equation}
    G_{i}^{(3)}=-J_{i}^{(3)}m_{i1}m_{i2}m_{i3}
\end{equation}
where $J_i^{(3)}$ can be $-1$ or $1$ with a minimal cost of $-1$.

\backmatter

\bmhead{Acknowledgements}
The numerical calculations in this paper have been done on the supercomputing system in the Supercomputing Center of Wuhan University. This work was supported in part by the Natural Science Foundation of Hubei Province, China under Grant 2024AFB784.

\section*{Declarations}


\begin{itemize}
\item \textbf{Data and materials availability} Public data have been cited in the references. Additional data and materials are available on reasonable request from the authors.
\item \textbf{Author contributions} H. W. initiated the project. H. Z. developed the code and performed the experiments. H. Z. and H. W. analyzed the data and contributed to the theoretical development. H.Z. and H.W. jointly wrote the manuscript. R. Z. contributed to FPGA code optimization. S. C. supervised the research. All authors contributed to the scientific discussions and provided feedback on the manuscript.
\item \textbf{Correspondence} Correspondence and requests for materials can be addressed to either H.Z.(HaiyangZ@whu.edu.cn) or H.W.(wanghao@whu.edu.cn).
\end{itemize}

\bibliography{sn-bibliography}


\begin{thebibliography}{26}
\ifx \bisbn   \undefined \def \bisbn  #1{ISBN #1}\fi
\ifx \binits  \undefined \def \binits#1{#1}\fi
\ifx \bauthor  \undefined \def \bauthor#1{#1}\fi
\ifx \batitle  \undefined \def \batitle#1{#1}\fi
\ifx \bjtitle  \undefined \def \bjtitle#1{#1}\fi
\ifx \bvolume  \undefined \def \bvolume#1{\textbf{#1}}\fi
\ifx \byear  \undefined \def \byear#1{#1}\fi
\ifx \bissue  \undefined \def \bissue#1{#1}\fi
\ifx \bfpage  \undefined \def \bfpage#1{#1}\fi
\ifx \blpage  \undefined \def \blpage #1{#1}\fi
\ifx \burl  \undefined \def \burl#1{\textsf{#1}}\fi
\ifx \doiurl  \undefined \def \doiurl#1{\url{https://doi.org/#1}}\fi
\ifx \betal  \undefined \def \betal{\textit{et al.}}\fi
\ifx \binstitute  \undefined \def \binstitute#1{#1}\fi
\ifx \binstitutionaled  \undefined \def \binstitutionaled#1{#1}\fi
\ifx \bctitle  \undefined \def \bctitle#1{#1}\fi
\ifx \beditor  \undefined \def \beditor#1{#1}\fi
\ifx \bpublisher  \undefined \def \bpublisher#1{#1}\fi
\ifx \bbtitle  \undefined \def \bbtitle#1{#1}\fi
\ifx \bedition  \undefined \def \bedition#1{#1}\fi
\ifx \bseriesno  \undefined \def \bseriesno#1{#1}\fi
\ifx \blocation  \undefined \def \blocation#1{#1}\fi
\ifx \bsertitle  \undefined \def \bsertitle#1{#1}\fi
\ifx \bsnm \undefined \def \bsnm#1{#1}\fi
\ifx \bsuffix \undefined \def \bsuffix#1{#1}\fi
\ifx \bparticle \undefined \def \bparticle#1{#1}\fi
\ifx \barticle \undefined \def \barticle#1{#1}\fi
\bibcommenthead
\ifx \bconfdate \undefined \def \bconfdate #1{#1}\fi
\ifx \botherref \undefined \def \botherref #1{#1}\fi
\ifx \url \undefined \def \url#1{\textsf{#1}}\fi
\ifx \bchapter \undefined \def \bchapter#1{#1}\fi
\ifx \bbook \undefined \def \bbook#1{#1}\fi
\ifx \bcomment \undefined \def \bcomment#1{#1}\fi
\ifx \oauthor \undefined \def \oauthor#1{#1}\fi
\ifx \citeauthoryear \undefined \def \citeauthoryear#1{#1}\fi
\ifx \endbibitem  \undefined \def \endbibitem {}\fi
\ifx \bconflocation  \undefined \def \bconflocation#1{#1}\fi
\ifx \arxivurl  \undefined \def \arxivurl#1{\textsf{#1}}\fi
\csname PreBibitemsHook\endcsname

\bibitem[\protect\citeauthoryear{Lenz}{1920}]{lenz1920beitrvsge}
\begin{barticle}
\bauthor{\bsnm{Lenz}, \binits{W.}}:
\batitle{Beitr{\v{s}}ge zum verst{\v{s}}ndnis der magnetischen eigenschaften in festen k{\v{s}}rpern}.
\bjtitle{Physikalische Z}
\bvolume{21}(\bissue{613-615}),
\bfpage{1}
(\byear{1920})
\end{barticle}
\endbibitem

\bibitem[\protect\citeauthoryear{Mohseni et~al.}{2022}]{mohseni2022ising}
\begin{barticle}
\bauthor{\bsnm{Mohseni}, \binits{N.}},
\bauthor{\bsnm{McMahon}, \binits{P.L.}},
\bauthor{\bsnm{Byrnes}, \binits{T.}}:
\batitle{Ising machines as hardware solvers of combinatorial optimization problems}.
\bjtitle{Nature Reviews Physics}
\bvolume{4}(\bissue{6}),
\bfpage{363}--\blpage{379}
(\byear{2022})
\end{barticle}
\endbibitem

\bibitem[\protect\citeauthoryear{Santoro et~al.}{2002}]{santoro2002theory}
\begin{barticle}
\bauthor{\bsnm{Santoro}, \binits{G.E.}},
\bauthor{\bsnm{Marton{\'a}k}, \binits{R.}},
\bauthor{\bsnm{Tosatti}, \binits{E.}},
\bauthor{\bsnm{Car}, \binits{R.}}:
\batitle{Theory of quantum annealing of an ising spin glass}.
\bjtitle{Science}
\bvolume{295}(\bissue{5564}),
\bfpage{2427}--\blpage{2430}
(\byear{2002})
\end{barticle}
\endbibitem

\bibitem[\protect\citeauthoryear{Hauke et~al.}{2020}]{hauke2020perspectives}
\begin{barticle}
\bauthor{\bsnm{Hauke}, \binits{P.}},
\bauthor{\bsnm{Katzgraber}, \binits{H.G.}},
\bauthor{\bsnm{Lechner}, \binits{W.}},
\bauthor{\bsnm{Nishimori}, \binits{H.}},
\bauthor{\bsnm{Oliver}, \binits{W.D.}}:
\batitle{Perspectives of quantum annealing: Methods and implementations}.
\bjtitle{Reports on Progress in Physics}
\bvolume{83}(\bissue{5}),
\bfpage{054401}
(\byear{2020})
\end{barticle}
\endbibitem

\bibitem[\protect\citeauthoryear{Inagaki et~al.}{2016}]{inagaki2016coherent}
\begin{barticle}
\bauthor{\bsnm{Inagaki}, \binits{T.}},
\bauthor{\bsnm{Haribara}, \binits{Y.}},
\bauthor{\bsnm{Igarashi}, \binits{K.}},
\bauthor{\bsnm{Sonobe}, \binits{T.}},
\bauthor{\bsnm{Tamate}, \binits{S.}},
\bauthor{\bsnm{Honjo}, \binits{T.}},
\bauthor{\bsnm{Marandi}, \binits{A.}},
\bauthor{\bsnm{McMahon}, \binits{P.L.}},
\bauthor{\bsnm{Umeki}, \binits{T.}},
\bauthor{\bsnm{Enbutsu}, \binits{K.}}, \betal:
\batitle{A coherent ising machine for 2000-node optimization problems}.
\bjtitle{Science}
\bvolume{354}(\bissue{6312}),
\bfpage{603}--\blpage{606}
(\byear{2016})
\end{barticle}
\endbibitem

\bibitem[\protect\citeauthoryear{Wang et~al.}{2013}]{wang2013coherent}
\begin{barticle}
\bauthor{\bsnm{Wang}, \binits{Z.}},
\bauthor{\bsnm{Marandi}, \binits{A.}},
\bauthor{\bsnm{Wen}, \binits{K.}},
\bauthor{\bsnm{Byer}, \binits{R.L.}},
\bauthor{\bsnm{Yamamoto}, \binits{Y.}}:
\batitle{Coherent ising machine based on degenerate optical parametric oscillators}.
\bjtitle{Physical Review A}
\bvolume{88}(\bissue{6}),
\bfpage{063853}
(\byear{2013})
\end{barticle}
\endbibitem

\bibitem[\protect\citeauthoryear{Su et~al.}{2022}]{su2022flexspin}
\begin{bchapter}
\bauthor{\bsnm{Su}, \binits{Y.}},
\bauthor{\bsnm{Kim}, \binits{T.T.-H.}},
\bauthor{\bsnm{Kim}, \binits{B.}}:
\bctitle{Flexspin: A scalable cmos ising machine with 256 flexible spin processing elements for solving complex combinatorial optimization problems}.
In: \bbtitle{2022 IEEE International Solid-State Circuits Conference (ISSCC)},
vol. \bseriesno{65},
pp. \bfpage{1}--\blpage{3}
(\byear{2022}).
\bcomment{IEEE}
\end{bchapter}
\endbibitem

\bibitem[\protect\citeauthoryear{Oku et~al.}{2019}]{oku2019fully}
\begin{barticle}
\bauthor{\bsnm{Oku}, \binits{D.}},
\bauthor{\bsnm{Terada}, \binits{K.}},
\bauthor{\bsnm{Hayashi}, \binits{M.}},
\bauthor{\bsnm{Yamaoka}, \binits{M.}},
\bauthor{\bsnm{Tanaka}, \binits{S.}},
\bauthor{\bsnm{Togawa}, \binits{N.}}:
\batitle{A fully-connected ising model embedding method and its evaluation for cmos annealing machines}.
\bjtitle{IEICE TRANSACTIONS on Information and Systems}
\bvolume{102}(\bissue{9}),
\bfpage{1696}--\blpage{1706}
(\byear{2019})
\end{barticle}
\endbibitem

\bibitem[\protect\citeauthoryear{Katzgraber et~al.}{2015}]{katzgraber2015good}
\begin{barticle}
\bauthor{\bsnm{Katzgraber}, \binits{H.G.}},
\bauthor{\bsnm{Hamze}, \binits{F.}},
\bauthor{\bsnm{Zhu}, \binits{Z.}},
\bauthor{\bsnm{Ochoa}, \binits{A.J.}},
\bauthor{\bsnm{Munoz-Bauza}, \binits{H.}}:
\batitle{Seeking quantum speedup through spin glasses: The good, the bad, and the ugly}.
\bjtitle{Physical Review X}
\bvolume{5}(\bissue{3}),
\bfpage{031026}
(\byear{2015})
\end{barticle}
\endbibitem

\bibitem[\protect\citeauthoryear{Dobrynin et~al.}{2024}]{dobrynin2024energy}
\begin{barticle}
\bauthor{\bsnm{Dobrynin}, \binits{D.}},
\bauthor{\bsnm{Renaudineau}, \binits{A.}},
\bauthor{\bsnm{Hizzani}, \binits{M.}},
\bauthor{\bsnm{Strukov}, \binits{D.}},
\bauthor{\bsnm{Mohseni}, \binits{M.}},
\bauthor{\bsnm{Strachan}, \binits{J.P.}}:
\batitle{Energy landscapes of combinatorial optimization in ising machines}.
\bjtitle{Physical Review E}
\bvolume{110}(\bissue{4}),
\bfpage{045308}
(\byear{2024})
\end{barticle}
\endbibitem

\bibitem[\protect\citeauthoryear{Commander}{2009}]{commander2009maximum}
\begin{botherref}
\oauthor{\bsnm{Commander}, \binits{C.W.}}:
Maximum cut problem, max-cut.
Encyclopedia of Optimization
\textbf{2}
(2009)
\end{botherref}
\endbibitem

\bibitem[\protect\citeauthoryear{Kowalsky et~al.}{2022}]{kowalsky20223}
\begin{barticle}
\bauthor{\bsnm{Kowalsky}, \binits{M.}},
\bauthor{\bsnm{Albash}, \binits{T.}},
\bauthor{\bsnm{Hen}, \binits{I.}},
\bauthor{\bsnm{Lidar}, \binits{D.A.}}:
\batitle{3-regular three-xorsat planted solutions benchmark of classical and quantum heuristic optimizers}.
\bjtitle{Quantum Science and Technology}
\bvolume{7}(\bissue{2}),
\bfpage{025008}
(\byear{2022})
\end{barticle}
\endbibitem

\bibitem[\protect\citeauthoryear{Hen}{2019}]{hen2019equation}
\begin{barticle}
\bauthor{\bsnm{Hen}, \binits{I.}}:
\batitle{Equation planting: a tool for benchmarking ising machines}.
\bjtitle{Physical Review Applied}
\bvolume{12}(\bissue{1}),
\bfpage{011003}
(\byear{2019})
\end{barticle}
\endbibitem

\bibitem[\protect\citeauthoryear{Peter}{1998}]{peter1998efficient}
\begin{botherref}
\oauthor{\bsnm{Peter}, \binits{A.}}:
Efficient shift registers, lfsr counters, and long pseudo-random sequence generators.
http://www. xilinx. com/bvdocs/appnotes/xapp052. pdf
(1998)
\end{botherref}
\endbibitem

\bibitem[\protect\citeauthoryear{Patel et~al.}{2020}]{patel2020ising}
\begin{botherref}
\oauthor{\bsnm{Patel}, \binits{S.}},
\oauthor{\bsnm{Chen}, \binits{L.}},
\oauthor{\bsnm{Canoza}, \binits{P.}},
\oauthor{\bsnm{Salahuddin}, \binits{S.}}:
Ising model optimization problems on a fpga accelerated restricted boltzmann machine.
arXiv preprint arXiv:2008.04436
(2020)
\end{botherref}
\endbibitem

\bibitem[\protect\citeauthoryear{Patel et~al.}{2022}]{patel2022logically}
\begin{barticle}
\bauthor{\bsnm{Patel}, \binits{S.}},
\bauthor{\bsnm{Canoza}, \binits{P.}},
\bauthor{\bsnm{Salahuddin}, \binits{S.}}:
\batitle{Logically synthesized and hardware-accelerated restricted boltzmann machines for combinatorial optimization and integer factorization}.
\bjtitle{Nature Electronics}
\bvolume{5}(\bissue{2}),
\bfpage{92}--\blpage{101}
(\byear{2022})
\end{barticle}
\endbibitem

\bibitem[\protect\citeauthoryear{Bremaud and Fields}{2001}]{bremaud2001markov}
\begin{barticle}
\bauthor{\bsnm{Bremaud}, \binits{P.}},
\bauthor{\bsnm{Fields}, \binits{M.C.G.}}:
\batitle{Monte carlo simulation, and queues (texts in applied mathematics 31)}.
\bjtitle{Statistics \& Decisions}
\bvolume{19},
\bfpage{315}--\blpage{330}
(\byear{2001})
\end{barticle}
\endbibitem

\bibitem[\protect\citeauthoryear{Wilmer et~al.}{2009}]{levin2017markov}
\begin{botherref}
\oauthor{\bsnm{Wilmer}, \binits{E.L.}},
\oauthor{\bsnm{Levin}, \binits{D.A.}},
\oauthor{\bsnm{Peres}, \binits{Y.}}:
Markov chains and mixing times.
American Mathematical Soc., Providence
\textbf{107}
(2009)
\end{botherref}
\endbibitem

\bibitem[\protect\citeauthoryear{Nikhar et~al.}{2024}]{nikhar2024all}
\begin{barticle}
\bauthor{\bsnm{Nikhar}, \binits{S.}},
\bauthor{\bsnm{Kannan}, \binits{S.}},
\bauthor{\bsnm{Aadit}, \binits{N.A.}},
\bauthor{\bsnm{Chowdhury}, \binits{S.}},
\bauthor{\bsnm{Camsari}, \binits{K.Y.}}:
\batitle{All-to-all reconfigurability with sparse and higher-order ising machines}.
\bjtitle{Nature Communications}
\bvolume{15}(\bissue{1}),
\bfpage{8977}
(\byear{2024})
\end{barticle}
\endbibitem

\bibitem[\protect\citeauthoryear{Hamerly et~al.}{2019}]{hamerly2019experimental}
\begin{barticle}
\bauthor{\bsnm{Hamerly}, \binits{R.}},
\bauthor{\bsnm{Inagaki}, \binits{T.}},
\bauthor{\bsnm{McMahon}, \binits{P.L.}},
\bauthor{\bsnm{Venturelli}, \binits{D.}},
\bauthor{\bsnm{Marandi}, \binits{A.}},
\bauthor{\bsnm{Onodera}, \binits{T.}},
\bauthor{\bsnm{Ng}, \binits{E.}},
\bauthor{\bsnm{Langrock}, \binits{C.}},
\bauthor{\bsnm{Inaba}, \binits{K.}},
\bauthor{\bsnm{Honjo}, \binits{T.}}, \betal:
\batitle{Experimental investigation of performance differences between coherent ising machines and a quantum annealer}.
\bjtitle{Science advances}
\bvolume{5}(\bissue{5}),
\bfpage{0823}
(\byear{2019})
\end{barticle}
\endbibitem

\bibitem[\protect\citeauthoryear{Krislock et~al.}{2017}]{krislock2017biqcrunch}
\begin{barticle}
\bauthor{\bsnm{Krislock}, \binits{N.}},
\bauthor{\bsnm{Malick}, \binits{J.}},
\bauthor{\bsnm{Roupin}, \binits{F.}}:
\batitle{Biqcrunch: A semidefinite branch-and-bound method for solving binary quadratic problems}.
\bjtitle{ACM Transactions on Mathematical Software (TOMS)}
\bvolume{43}(\bissue{4}),
\bfpage{1}--\blpage{23}
(\byear{2017})
\end{barticle}
\endbibitem

\bibitem[\protect\citeauthoryear{Wiegele}{2007}]{wiegele2007biq}
\begin{barticle}
\bauthor{\bsnm{Wiegele}, \binits{A.}}:
\batitle{Biq mac library—a collection of max-cut and quadratic 0-1 programming instances of medium size}.
\bjtitle{Preprint}
\bvolume{51},
\bfpage{112}--\blpage{127}
(\byear{2007})
\end{barticle}
\endbibitem

\bibitem[\protect\citeauthoryear{Helmberg and Rendl}{2000}]{helmberg2000spectral}
\begin{barticle}
\bauthor{\bsnm{Helmberg}, \binits{C.}},
\bauthor{\bsnm{Rendl}, \binits{F.}}:
\batitle{A spectral bundle method for semidefinite programming}.
\bjtitle{SIAM Journal on Optimization}
\bvolume{10}(\bissue{3}),
\bfpage{673}--\blpage{696}
(\byear{2000})
\end{barticle}
\endbibitem

\bibitem[\protect\citeauthoryear{Rinaldi}{1995}]{rudy}
\begin{botherref}
\oauthor{\bsnm{Rinaldi}, \binits{G.}}:
{Rudy: A Rudimental Graph Generator}.
GitHub repository
(1995).
\url{https://github.com/g-rinaldi/rudy}
\end{botherref}
\endbibitem

\bibitem[\protect\citeauthoryear{Isakov et~al.}{2015}]{isakov2015optimised}
\begin{barticle}
\bauthor{\bsnm{Isakov}, \binits{S.V.}},
\bauthor{\bsnm{Zintchenko}, \binits{I.N.}},
\bauthor{\bsnm{R{\o}nnow}, \binits{T.F.}},
\bauthor{\bsnm{Troyer}, \binits{M.}}:
\batitle{Optimised simulated annealing for ising spin glasses}.
\bjtitle{Computer Physics Communications}
\bvolume{192},
\bfpage{265}--\blpage{271}
(\byear{2015})
\end{barticle}
\endbibitem

\bibitem[\protect\citeauthoryear{Sajeeb et~al.}{2025}]{sajeeb2025scalable}
\begin{botherref}
\oauthor{\bsnm{Sajeeb}, \binits{M.}},
\oauthor{\bsnm{Aadit}, \binits{N.A.}},
\oauthor{\bsnm{Wu}, \binits{T.}},
\oauthor{\bsnm{Smith}, \binits{C.}},
\oauthor{\bsnm{Chinmay}, \binits{D.}},
\oauthor{\bsnm{Raut}, \binits{A.}},
\oauthor{\bsnm{Camsari}, \binits{K.Y.}},
\oauthor{\bsnm{Delacour}, \binits{C.}},
\oauthor{\bsnm{Srimani}, \binits{T.}}:
Scalable connectivity for ising machines: Dense to sparse.
arXiv preprint arXiv:2503.01177
(2025)
\end{botherref}
\endbibitem

\end{thebibliography}

\end{document}